\documentstyle[12pt]{article}

\begin{document}
\def\be{\begin{equation}}
\def\ee{\end{equation}}
\def\Chr#1#2{\left\{^{#1}_{#2}\right\}}
\def\half{\frac{1}{2}}
\def\rg{\sqrt{-g}}
\def\A{{\cal A}}
\def\a{\alpha}
\def\t{\tau}
\def\b{\beta}
\def\m{\mu}
\def\n{\nu}
\def\k{\kappa}
\def\g{\gamma}
\def\G{\Gamma}
\def\e{\eta}
\def\E{\varepsilon}
\def\l{\lambda}
\def\L{{\cal L}}
\def\Rt{\tild{R}}
\def\Ne{$\nu_e\;$}
\def\Nm{$\nu_\mu\;$}

\begin{titlepage}
\vspace*{10mm}
\begin{center} {\Large \bf Neutrino oscillation in a space-time with torsion} \\
\vskip 10mm
\centerline {\bf
M. Alimohammadi$ ^{a,c}$ \footnote {e-mail:alimohmd@netware2.ipm.ac.ir},
A. Shariati$ ^{b,c}$}
\vskip 1cm
{\it $^a$ Department of Physics, University of Teheran, North Karegar,} \\
{\it Tehran, Iran }\\
{\it $^b$ Institute for Advanced Studies in Basic Physics , P.O.Box 159 ,}\\
{\it  Gava Zang , Zanjan 45195 , Iran }\\
{\it $^c$ Institute for Studies in Theoretical Physics and Mathematics,}\\
{\it P.O.Box 19395-5531, Tehran, Iran}\\
\end{center}

\vskip 2cm
\begin{abstract}
\noindent
Using the Einstein-Cartan-Dirac theory, we study the effect of torsion
on neutrino oscillation. We see that torsion cannot induce
neutrino oscillation, but affects it whenever oscillation exists for
other reasons. We show that the torsion effect on neutrino oscillation
is as important as the neutrino mass effect, whenever
the ratio of neutrino number density to neutrino energy is
$\sim 10^{69}$ cm$^{-3}$ /eV, or the number density of the matter is
$\sim 10^{69}$ cm$^{-3}$.
\end{abstract}

\end{titlepage}
\newpage
\section{Introduction}
Results from several solar neutrino experiments, utilizing differnent
detection techniques, consistently show a discrepancy between the
measured \Ne flux from the sun and the \Ne flux predicted by various
solar models. The origin of this solar neutrino deficit is not yet
well knwon. A possible solution is neutrino flavour oscillations,
which was first suggested by Pontecorvo \cite{Pon} and then
Maki {\em et al.} \cite{Maki}.

Several mechanisms for neutrino oscillations have been proposed
(see for example \cite{KP}). One mechanism, for example, assumes
that neutrinos have non-equal masses, and that the neutrino mass
eigenstates are not the weak interaction eigenstates.
The most famous version of this type of solution is the
Mikheyev-Smirnov-Wolfenstein (MSW) mechanism \cite{Wolf,MS}.
There have been also proposed some alternative mechaninsms,
relating neutrino oscillations to gravitaional effects; {\it e.g.}
to violation of equivalence principle \cite{Gasp}, or to violation of
the Lorentz group symmetry \cite{CG}. There have been also some efforts
to study the effect of (possible) torsion of space-time on
neutrino oscillations \cite{SG2,SS}.
In ref. \cite{SG2}, the authors have studied this effect, by assuming
that the torsion eigenstates, {\it i.e.} the eigenstate of the
interaction part of the hamiltonian $H_T$, are different from
the weak interaction eigenstates. In the case that the two eigenstates
are the same ($\theta_T = 0$, in their language), their results
coincide with the ordinary vacuum oscillation and the effect of
torsion does not appear. Here, we study this problem from a different
point of view. First, we take the torsion eigenstates the same as
the weak interaction ones. Second, our procedure is different from
\cite{SG2} and based on the technique that was introduced in \cite{CF}.
We use this technique in the Einstein-Cartan-Dirac theory.

The structure of the paper is as follows.
In section 2, we  briefly discuss the
Einstein-Cartan-Dirac theory, in which a Dirac field couples to the metric
and torsion of the space--time. We see that in this theory, the torsion is
completely determined by a pseudo--vector $\A_\m$, and the equation of
motion fixes it to be proportional to the axial current of the Dirac particle.
In section 3, we study the effect of torsion on neutrino oscillation, and
show that although the torsion can not induce neutrino oscillation, but affects
it whenever oscillation exists for other reasons.
We find some lower bounds on physical quantities such that the
oscillation induced by torsion be of the same order as the oscillation
induced by neutrino mass.

\section{Breif review of Einstein-Cartan-Dirac theory}
The differential geometry of a four dimensional
manifold $U^4$ is determined by two objects;
the Riemannian metric $g_{\m\n}$, and the connection $\G^\m_{\a\b}$.
Conceptualy these two objects are completely independent.
The metric determines the inner product of vectors at each point,
enabling us to define arclengths and thus distances. The connection
determines the parallel transport and thus the covariant derivative of
tensor and spinor fields. The connection is said to be compatible
with metric, if the parallel transport of vectors does not change
both the length of vectors and the angle  between them.

The most general connection, compatible with the Riemannian metric, is
\be \G^\a_{\m\n} := \Chr{\a}{\m\n} + K^\a_{\m\n}, \ee
where
\be \Chr{\a}{\m\n} = \half g^{\a\b} \left( -g_{\m\n,\b} + g_{\b\m,\n}
+ g_{\n\b,\m} \right), \ee
is the usual Christoffel symbole, and $K^\a_{\m\n}$ is a rank 3 tensor,
called contorsion.
The only restriction on $K$ is that, when its upper index
is lowered with the metric, it has the following symmetry
property
\be K_{\a\m\b} = - K_{\b\m\a}. \ee
It therefore follows that, in a $d$-dimensional space-time,
it has $d^2(d-1)/2$ independent components.

Contorsion is related to the torsion tensor as follows
\be K^\a_{\m\n} := \half g^{\a\b} \left( T_{\b\m\n} + T_{\m\b\n}
+ T_{\n\b\m} \right), \ee
and the torsion $T^\a_{\m\n}$ itself is the antisymmetric
part of the connection
\be T^\a_{\m\n} = \G^\a_{\m\n} - \G^\a_{\n\m}. \ee
It follows that the differential geometry of $U^4$  is determined
by two tensor fields: 1) the metric tensor $g_{\m\n}$ and,
2) the contorsion tensor $K^\a_{\m\n}$, or equivalently
the torsion tensor $T^\a_{\m\n}$.

In general relativity, the space-time is considered to be torsion free,
a priori, while in Eistein-Cartan theory this is not the case.
Here, we quickly reveiw this latter theory.

As usual, the curvature tensor is defined as
\be
R^\k_{\l\m\n} := \G^\k_{\n\l,\m} - \G^\k_{\m\l,\n}
 + \G^\e_{\n\l} \G^\k_{\m\e} - \G^\e_{\m\l} \G^\k_{\n\e}.
\ee
Let $V^4$ denote the same manifold with the same Riemannian metric, but
with vanishing torsion. The Riemann curvature of $V^4$ is
\be
^0R^\k_{\;\l\m\n} := \Chr{\k}{\n\l}_{,\m} - \Chr{\k}{\m\l}_{,\n}
 + \Chr{\e}{\n\l} \Chr{\k}{\m\e} - \Chr{\e}{\m\l} \Chr{\k}{\n\e}.
\ee
Defining
\be \tau_\m := g^{\a\b} K_{\a\b\m}, \ee
\be \A^\sigma := \frac{1}{3} \E^{\sigma\a\m\n}
K_{\a\m\n}, \ee
the contorsion can be written as
\be K_{\a\m\n} = \frac{1}{3} \left( g_{\a\m}\tau_\n - g_{\n\m}\tau_\a \right)
+ \frac{1}{2} \A^\sigma \E_{\sigma\a\m\n} + U_{\a\m\n}. \ee
This expression is simply the decomposition of contorsion tensor. Here
$U_{\a\m\n}$ is in fact defined by the above equation, and has the following
properties
\be
U_{\a\m\n} = -U_{\n\m\a}, \;\; g^{\a\m}U_{\a\m\n} = 0,
\;\; \E^{\sigma\a\m\n}U_{\a\m\n} =0. \ee
$\E_{\a\k\m\n}$ is the totally antisymmetric pseudo-tensor of
rank 4.
Now, It can be shown that the scalar curvature is
\be \label{ECaction}
R = ^0R - \frac{2}{\sqrt{g}} \partial_\k(\sqrt{g}\tau^\k) +
\left(-\frac{1}{3} \tau^2 + \frac{3}{2}\A^2 + U_{\a\m\n}U^{\m\n\a} \right),
\ee
where $\sqrt{g}=[-{\rm det }(g_{\m\n})]^{1/2}$.
Einstein-Cartan theory is a theory of gravitation in which the space-time
is a manifold with torsion ({\it i.e.}, in our notation a $U^4$). The action
functional of the Eistein-Cartan theory is
\be I_{{\rm EC}} := - \frac{c^3}{16\pi G}
\int d^4 x \sqrt{g} R= \ee
\be - \frac{c^3}{16\pi G} \int d^4 x \left( \sqrt{g}
\;^0R - 2 \partial_\k (\sqrt{g}\tau^\k) + \sqrt{g}
\left(- \frac{1}{3} \tau^\k \tau_\k + \frac{3}{2} \A^\k \A_\k
+ U_{\a\m\n}U^{\m\a\n}
\right)
\right).
\ee
An important feature of this action is that the contorsion contribute
in a total derivative plus an algebraic expression. Therefore,
in Einstein-Cartan
theory, the {\it equation of motion} of the torsin field is simply an algebraic
equation
\be \A^\k = 0, \;\; \tau^\k = 0, \;\; U_{\a\m\n}=0. \ee
In other words, in the absence of matter, the Einstein-Cartan action
implies that there is no torsion.

Now let us couple this to a spin--$1/2$ field. The resulting theory
is known as the Einstein-Cartan-Dirac (ECD) theory.
The action of this theory is
\be \label{IECD} I_{{\rm ECD}} := I_{{\rm EC}} + I_{{\rm D}}, \ee
where
\be \label{ID} I_{{\rm D}} := \int d^4 x \sqrt{g} (-\hbar)
 \bar{\psi}\left(e_a^\m \g^a (\partial_\m + \Gamma_\m )
+ \frac{m c}{\hbar} \right) \psi . \ee
Here $a$ is a tetrad index and $\m$ is a coordinate index.
The coupling of metric and torsion to the Dirac field
is through the covariant derivative $D_\m : = \partial_\m + \G_\m$,
in which $\G_\m$ is the spin connection,
\be \G_\m :=-\frac{i}{8} [\g^a, \g^b] e_{a}^{\n} e_{b\n\vert\m}\; . \ee
Here a vertical line $\vert$ means the covariant derivative on $U^4$
\be e_{b\m\vert\n} := e_{b\m,\n} - \G^\l_{\m\n} e_{b\l}
= e_{b\m,\n} - \Chr{\l}{\m\n} e_{b\l} - K^\l_{\m\n} e_{b\l}.
\ee
We can write this as follows
\be e_{b\m\vert\n} := e_{b\m;\n}
- K^\l_{\m\n} e_{b\l},
\ee
where
\be \label{vs}
e_{b\m;\n} := e_{b\m,\n} - \Chr{\l}{\m\n} e_{b\l},
\ee
is the covariant derivative on $V^4$.

It can be easily seen that the effect of torsion in (\ref{ID})
is to add  to  the usual Dirac action
in a curved torsionless space-time ($V^4$), the following interaction term
\be
\int d^4x \sqrt{g} (6\hbar \A_\m J^\m_5), \ee
where $J_5^\m=-i{\bar \psi}\g^\m\g^5\psi$ is the axial current
of the Dirac field. Now, taking variation of (\ref{IECD}) with respect to
$\A^\m$, $\tau^\m$, and $U_{\a\m\n}$, leads to the following
equations of motion
\be \tau^\m = 0, \qquad U_{\a\m\n} =0, \ee
\be \label{hasan}
\A^\m = \frac{96\pi}{3} \frac{G\hbar}{c^3} J^\m_5. \ee
In summary, in ECD theory the torsion pseudo-vector is proportional to
the axial current of the fermion field, and this current is the source
of torsion.
The proportionality constant
is the square of Planck length.

\section{ Effect of torsion on neutrino oscillation}
Now everything is ready to study the effect of torsion on neutrino oscillation.
We follow the same procedure that was introduced in \cite{CF}, in which the
neutrino oscillation in curved background was studied. If we ignore the
background matter effect, the Dirac field equation of motion is
\be \label{30}
\left[\g^a e_a^\m \left( \partial_\m + \G_\m \right) + \frac{mc}{\hbar}
 \right] \psi
= 0, \ee
where the torsion contribution comes from $\G_\m$ term, via eqs.(18) and (19).
Using the identity
\be
\g^a [\g^b, \g^c] = 2 \e^{ab} \g^c - 2 \e^{ac} \g^b - 2i\E^{dabc} \g_5 \g_d,
\ee
one can show that the only nonvanishing contribution from spin connection is
\be
\g^a e_a^\m \G_\m = \g^a e_a^\m
\left\{
 i A_\m \left[
  - \frac{1}{2\sqrt{g}} \g_5
 \right]
\right\},
\ee
where
\be \label{31}
A^\m = A_G^\m + A_T^\m.
\ee
Here
\be
A_G^\m = \frac{1}{4} \sqrt{g} e_a^\m \E^{abcd} (e_{b\n,\sigma} -
  e_{b\sigma,\n}) e_c^\n e_d^\sigma ,
\ee
is the contribution of the Christoffel symbole, which is present in
both $V^4$ and $U^4$ and was drived in \cite{CF}, and
\be \label{32}
A_T^\m =  - 6 \sqrt{g} \A^\m ,
\ee
is the contribution of the torsion filed $\A^\m$, which is present
only in $U^4$. Now, a look at eqs.(\ref{hasan}), (\ref{30}), (\ref{31}),
and (\ref{32})
shows that in Einstein-Cartan-Dirac theory,
there is a self interaction among
the fermion field, because of torsion. This interaction, however, is
very small, because of the coefficient in (\ref{hasan})\footnote {There is another
point of view about the coupling constant that appear in spin--torsion interaction
[9,11-13]. The torsionic contact interaction Lagrangian between two spin half
particles is formally identical to the weak interaction Lagrangian and may be
written in the $(V-A)$ form, if at least one of the two fermions is massles.
This suggest that the spin torsion coupling constant $G_T$, be also identified
with the weak interaction Fermi constant. This suggest $G_T/G \approx 10^{31}$.}.

As was argued in \cite{CF}, we can add a term proportional to the identity to
eq.(27), to group it with term arising from matter effects. The result is
\be
\g^a e^\m_a \G_\m =\g^a e^\m_a (iA_\m {\cal P}_L),
\ee
where ${\cal P}_L$ is the left--handed projection operator. Putting (31) in
(25), we see that the momentum operator $P_\m$, used in neutrino oscillation
calculation, can be computed from following mass shell condition
\be
(P_\m +\hbar A_\m'{\cal P}_L)(P^\m +\hbar A'^\m {\cal P}_L)=-M_f^2 c^2,
\ee
where now $A_\m'$ representing both spin connection $A_\m$ and matter ($A_{f\m}$)
contribution
\be
A_\m'=A_\m +A_{f\m},
\ee
with \cite{CF}
\be
A_{f\m}= \left( \begin{array}  {cc} -{\sqrt 2}\frac{G_F}{\hbar c^2}
N_e^\m &0 \\ 0&0
\end{array} \right).
\ee
$M_f^2$ is the vacuum mass matrix in flavour basis
\be
M_f^2=U
\left( \begin{array}  {cc} m_1^2 &0 \\ 0&m_2^2
\end{array} \right) U^\dagger ,
\ee
with
\be
U=\left( \begin{array}  {cc} {\rm cos} \theta &{\rm sin} \theta \\ -{\rm sin} \theta & {\rm cos} \theta
\end{array} \right) .
\ee
$A_{f\m}$ is the potential, drived in the context of MSW effect, $N_e=n_eu^\m$
is the number current of the electron fluid; $n_e$ is the electron density in
the fluid rest frame, and $u^\m$ is the fluid's four velocity. $G_F$ is Fermi
constant.

Now for a general trajectory with affine parameter $ \lambda$
($x^\m =x^\m (\lambda )$), and for relativistic neutrinos, ignoring
terms of $O(A'^2)$ and $O(A'M_f)$, one finds that
column vector of flavour amplitudes
\be
\chi (\lambda)=
\left( \begin{array}  {c} <\n_e |\psi (\lambda)> \\ <\n_\m |\psi (\lambda)>
\end{array} \right) ,
\ee
satisfies in the following differential equation \cite{CF}
\be
i{d\chi \over d\lambda}=({M_f^2 c^2 \over 2} + \hbar p.A'{\cal P}_L )\chi ,
\ee
where $p^\m =dx^\m /d\lambda$ is the tangent vector to the null world line.
In the above relation $P^0=p^0$ and $P^i=(1-\epsilon )p^i$ ($\epsilon <<1 $).
In this way one can calculate the effect of torsion, through eqs.(33) and (28),
on neutrino oscillation. As in the case that was considered in \cite{CF}, the
total gravitational contribution $A_\m$ is proportional to identity matrix in
flavour space, and can not induce neutrino oscillation on its own, but affects
it when there are other off-diagonal terms.

To evaluate the order of magnitude of the effect of torsion on oscillation,
let us consider a case with only the mass and torsion terms. We want
to study the conditions under which the effect of torsion is of the
same order as the mass effect, {\it i.e.}:
\be \label{yek}
\frac{1}{2} m_\n^2 c^2 \sim 6 \sqrt{g} \frac{96\pi}{3} \frac{G\hbar}{c^3}
p_\n \cdot J_5. \ee
For a spin--1/2 particle with $S_z = \hbar/2$, momentum
$p_s^\m =( E_s/c, 0,0, p_s)$, and density $\rho_s$ we have
\be \label{do}
J_5^\m = \rho_s (\frac{p_s c}{E_s},0,0,1). \ee
Also, noting that $P^2 = - m_\n^2 c^2$ and $p_\n^2 = 0$, up to order
$\epsilon$ we have
\be \label{se}
p_\n^\m = (\frac{E_\n}{c}, 0, 0, (1 + \frac{m_\n^2 c^2}{2P_\n^2})P_\n).\ee
If the source of torsion is also neutrino, $p_s = P_\n$ and
$E_s = E_\n$, eq. (\ref{yek}) results
\be \frac{\rho_\n}{E_\n} \sim 10^{69} \, \frac{{\rm cm}^{-3}}{{\rm eV}}. \ee
This shows that torsion can affect the neutrino oscillation whenever its
number density is very large or its energy is very low. If the source of
torsion are some other spin--1/2 particles at rest, such as electrons or neutrons,
eq. (\ref{yek}) restricts the matter number density as follows
\be \rho_{{\rm matter}} \sim 10^{69} \, {\rm cm}^{-3}. \ee
\\ {\bf Acknowledgement} \\ M. Alimohammadi would like to
thank the research council of Tehran University, for partial
financial support.

\end{document}